\documentclass{llncs}
\usepackage{makeidx}  
\usepackage{algorithm}
\usepackage{algorithmicx}
\usepackage{algpseudocode}
\usepackage{amsmath}
\usepackage{amssymb}
\usepackage{url}
\usepackage{graphicx}
\title{Approximation Algorithms for Steiner Tree
Problems Based on Universal Solution Frameworks\thanks{Research was supported by  the ERC StG project PAAl no.
259515.}}

\begin{document}

\author{Krzysztof Ciebiera \and Piotr Godlewski \and Piotr Sankowski \and Piotr
Wygocki\\ \email{ciebie,pgodlewski,sank,wygos@mimuw.edu.pl}}

\institute{Institute of Informatics, University of Warsaw, Poland}

\maketitle              

\begin{abstract}
This paper summarizes the work on implementing few solutions
for the Steiner Tree problem which we undertook in the PAAL project. The
main focus of the project is the development of generic implementations
of approximation algorithms together with universal solution frameworks.
In particular, we have implemented Zelikovsky 11/6-approximation using local search
framework, and 1.39-approximation by Byrka {\it et al.} using iterative rounding framework.
These two algorithms are experimentally compared with greedy 2-approximation, with exact but exponential
time Dreyfus-Wagner algorithm, as well as with results given by a state-of-the-art local search techniques by Uchoa and Werneck.
The results of this paper are twofold. On one hand, we demonstrate that high level algorithmic
concepts can be designed and efficiently used in C++. On the other hand, we show that
the above algorithms with good theoretical guarantees, give decent results in practice, but are inferior to state-of-the-art heuristical
approaches.
\end{abstract}
\newpage
\pagenumbering{arabic}
\pagestyle{plain}

\section{Introduction}
Nowadays, working on state-of-the-art approximation algorithms requires the knowledge of many high-level tools
and concepts. For example, a very successful line of research in approximation algorithms was based on
iterative rounding idea~\cite{laubook}, which lead to an approximation algorithm for the
Steiner tree problem with the best theoretical approximation guarantee~\cite{byrka}. This means that
implementing and testing such algorithms in practice is even harder, because one
not only needs to understand these high level concepts but also is required to implement them. In our PAAL project\footnote{The Practical Approximation Algorithms Library (PAAL) is a header-only, generic library consisting of approximation algorithms, data structures and
several complete solutions for various optimization problems, implemented in C++11 available at \url{http://paal.mimuw.edu.pl}.} we have undertook a task to provide C++11 implementations of such high level tools including the iterative
rounding, and local search frameworks. These frameworks have been implemented having the following
design considerations in mind:
\begin{description}
\item[Easiness to use] It should be always possible to build the approximation
algorithm by implementing only functions, i.e., no definition of classes are needed.
\item[Minimalism] Our design minimizes number of functions one
needs to write in order to achieve required results. Depending on
selected optimization method programmer needs to provide only
functions required for this method.
\item[Speed] Our library exploits benefits of static polymorphism supporting programs in both object-oriented and
functional style. It enables compiler to use more sophisticated code optimization methods, function
inlining, loop unrolling, branch prediction, etc. \cite{Driesen:1996}.
\item[Loose coupling] Library elements, as much as possible, do not depend on each other. It is possible to change behaviour of the
solver by changing only one of its elements.
\item[Extensibility] One can add new solution strategies, e.g., search strategies, without modifying other elements of the framework.
\end{description}
In the case of local search these design assumption led to a framework, that
on one hand, requires no class and much less function definitions
when comparing it to other existing libraries like: Paradiso, Metslib, or Easylocal~\cite{LS}.
On the other hand, the running time of our implementations, when counting only the time spend in
the library functions, is at least 3 time smaller~\cite{LS}.

The aim of this paper is to report on the PAAL implementations
of solutions for the Steiner tree problem. In this problem we are given an undirected graph $G = (V, E)$,
with edge costs (weights) $c : E \to \mathbb{Q}_+$, and a subset of nodes $T \subseteq V$ (terminals), the {\em Steiner
Tree }problem asks for a tree $S$ spanning the terminals, of minimum cost $c(S)
= \sum_{e\in S} c(e)$. Note that $S$ might contain some other nodes, besides
the terminals (Steiner nodes).

Steiner Tree problem is NP-hard and even APX-hard to approximate~\cite{Chlebik2008207}, i.e.,
approximating it better then $96/95=1.0105\ldots$ is NP-hard.
During recent years it has become a benchmark problem in approximation algorithms study. Three
main techniques have been used to design approximation algorithms with good theoretical guarantees for this problem:
\begin{itemize}
\item greedy approach gives a $2$-approximation~\cite{gilbert1968steiner},
\item Zelikovsky introduced local search to obtain an $11/6$-approximation~\cite{zel},
\item the best $1.39$-approximation was proposed by Byrka {\it et al.}~\cite{byrka} and
uses iterative rounding.\footnote{For the full history of the theoretical studies of this problem please see~\cite{byrka}.}
\end{itemize}
We have implement the above 3 algorithms together with an exponential time exact algorithm
given in~\cite{dre}. The algorithm of Zelikovsky was implemented using our local search framework,
whereas the algorithm by Byrka {\it et al.} was implemented using our iterative rounding framework.
More details on the implementations and on how these frameworks aided us are given in the following sections of this paper.
We have decided to compare our algorithms with results given by the state-of-the-art local search solution
by Uchoa and Werneck~\cite{wer}.\footnote{We would like to thank Renato Werneck for giving us these results, so
we did not have to reimplement their solution.} For completeness of this paper we give some details
of this solution in Section~\ref{sec:local-search}. The final section of this paper gives the result
of our experimental study. A priori we were suspecting that the iterative rounding algorithm could deliver
comparable results to the state-of-the-art heuristic solutions, as it was the case for the Minimum Bounded-Degree Spanning Tree (MBDST) problem~\cite{BernathCGS14}. However, the experiments show that this is not the case. On SteinLib~\cite{stlib} instances iterative rounding has an average approximation rato of 1.029, whereas the solution from~\cite{wer} gives 1.01 on average. In other words, the additive error of our solution is
on average 3 times higher. There are only few test cases, where iterative rounding found better answers. However, we note that the paper~\cite{wer} considers 12 different local search algorithms, and only the best one of them visibly outperforms our iterative rounding implementation.

It appears that the main weakness of this iterative rounding solution is the need to generate all
$k$-terminal components (i.e., $k$-terminal subtrees) from which an approximate Steiner tree can be build. On one hand, it seems that one really needs  larger values of $k$ to guarantee good approximation ratio. On the other hand, the generation of all such components is a bottleneck in the running time. When this procedure was implemented following exactly the description in~\cite{byrka} it took 98\% of the running time needed by the algorithm. We have came up with several optimizations for this procedure, but even using them it still consumes 80\% of the running time. We note that our implementation precedes the simplifications of Goemans {\it et al.}~\cite{Goemans} to the algorithm of Byrka {\it et al.}~\cite{byrka}. However, these improvements are unlikely to have practical impact as the above bottleneck is still present there. Although, the implemented algorithms do not outperform state-of-the-art heuristics, our implementations have demonstrated that high level approximation algorithms can be implemented in an efficient and extendable way. As we have
already mentioned our local search framework is easier and faster then alternative solutions, and PAAL library contains local search solutions for the following problems: traveling salesman problem, facility location, $k$-median, and capacitated facility location. On the other, our iterative rounding framework allowed us to easily implement solutions to the following additional problems: bounded-degree minimum spanning tree, generalised assignment,  Steiner network, and tree augmentation.

The paper is composed as follows. The following four sections give the implementation details of greedy, Zelikovsky, Dreyfus-Wagner, and iterative rounding algorithms. Next, some details of the local search heuristics are given. Finally, Section~\ref{sec:exper} contains the description of our experiments.

\section{Greedy 2-approximation}
Let $G^*$ be the metric closure of the graph $G$, and given a weighted graph $H$ we denote by MST$(H)$ the minimum spanning
tree of $H$. It is well known that the minimum spanning tree of a metric closure of the graph restricted
to terminals $T$ (i.e., MST$(G^*[T])$) is a 2-approximation of the Steiner Tree
problem~\cite{vazirani}. The time complexity of a naive implementation of the above algorithm equals
$O(|T||E|\log(|V|))$, i.e., when one computes the distances from each terminal.
PAAL implementation of this algorithm performs one run ofthe Dijkstra's algorithm starting from all terminals at once
(see~\cite{Mehlhorn1988125} for more details).  This optimization helped us to reduce
the time complexity of the algorithm to $O(|E|\log(|V|))$ (see~\cite{WygosGreedyStein2014} for the implementation).

\section{11/6-approximation}

The second algorithm implemented in PAAL is an $11/6$-approximation by
Zelikovsky~\cite{zel}. We implemented the faster $O(|V|\cdot|E|+|T|^4)$ time
complexity version of this algorithm. The algorithm is in a form
of a local search, so we use PAAL's Local Search framework for the implementation.
In particular we use the Hill Climbing primitive, i.e., we start with some solution and improve it as long as it
is possible. PAAL provides framework for Hill Climbing consisting of
three primitives (components):
\begin{itemize}
    \item \emph{State} -- current solution,
    \item \emph{Neighbourhood} -- list of moves that can be applied to a state,
    \item \emph{Gain} -- difference between the value of a state after
        and before applying a move.
\end{itemize}

We present the outline of our implementation of Zelikovsky's algorithm.
First the algorithm builds some initial data structures:
\begin{itemize}
	\item Minimum Spanning Tree on the set of terminals,
	\item Voronoi regions of terminals: sets of Steiner vertices which are
		closer to a given terminal than to any other terminal,
	\item centers of terminal triples: for each triple of terminals we find
		its center, that is a Steiner vertex which minimizes the sum of distances to
		the terminals in the triple.
\end{itemize}

At each iteration the algorithm builds recursively a save matrix $M$ (as defined in
\cite{zel}) with rows and columns labeled by current terminals (some terminals
are contracted during the algorithm). For any given pair of terminals
$T_1, T_2$, the element of the save matrix $M[T_1, T_2]$, contains the cost
of the most expensive edge on the cheapest path from $T_1$ to $T_2$.
Next, by using Hill Climbing method, the algorithm iteratively
improves the tree by adding new Steiner points to it. Each
added point is a center of terminal triple and after adding it algortithm
contracts the triple by setting costs of edges between tripple's vertices
to $0$. The pseudocode of the algorithm
is given as Algorithm~\ref{alg:zel}.

\begin{algorithm}
    \begin{algorithmic}
        \State $tree \gets$ MST($G^*[T]$)
        \State $selected\_nonterminals \gets \emptyset$
        \State find Voronoi regions of all $terminal$s
        \For{each $triple \in$ triples of terminals}
        \State $center(triple) \gets$ center of $triple$ (as defined before)
        \State $cost(triple) \gets$ sum of distances from the center to triple vertices
        \EndFor
        \Loop
        \State $save \gets$ save matrix of the $tree$
        \State $move \gets$ triple which maximizes:
        $$gain \gets \max\limits_{e\in triple}save(e) +
        \min\limits_{e\in triple}save(e) - cost(triple)$$
        \If {$gain \le 0$}
        exit loop
        \Else
        \State contract $move$
        \State $selected\_nonterminals \gets selected\_nonterminals + center(move)$
        \EndIf
        \EndLoop
        \State \Return $MST(G^*[T \cup selected\_nonterminals])$
    \end{algorithmic}
    \caption{Pseudocode of Zelikovsky algorithm}
    \label{alg:zel}
\end{algorithm}

Graph operations were implemented using Boost Graph Library~\cite{bgl}.
Algorithm for computing Voronoi regions is implemented as a part of PAAL.
Calculation of the $save$ matrix is implemented recursively as it was
presented in the original paper. Full C++ code can be found at~\cite{WygosZel2014}.

\section{Dreyfus-Wagner Algorithm}
The Dreyfus-Wagner algorithm~\cite{dre} finds an optimum solution to the Steiner Tree
problem in exponential time (with respect to the number of terminals): $O(3^{|T|}∗|V|+2^{|T|}∗|V|^2)$. It will
be also used to solve subproblems in the Iterative Rounding algorithm.

Our implementation is a straightforward recursive implementation
of the Dreyfus and Wagner dynamic programming methods. For $X\subseteq T$ and
$v\in V\setminus X$ we define $C(v,X)$ as the minimum cost of the Steiner tree spanning $X\cup\{v\}$
and $B(v,X)$ as the minimum cost of the Steiner tree spanning $X\cup\{v\}$, where $v$
has degree at least two. The Dreyfus-Wagner algorithm is based on the following
recursive formulas. The first formula comes from the fact, that given an optimal
Steiner tree spanning $X\cup\{v\}$ in which the degree of $v$ is at least two,
we can split the tree at $v$ into two subtrees: one spanning $Y\cup\{v\}$ and one
spanning $(X\setminus Y)\cup\{v\}$, hence:
\begin{equation}
B(v,X) = \min_{\emptyset\subset Y\subset X} \{C(v,Y)+C(v,X\setminus Y)\}.
\end{equation}
To get the second formula, let us consider the optimal Steiner tree spanning $X\cup\{v\}$,
in which the degree of $v$ is 1. In such case, the tree path form $v$ leads either to
a vertex $u\in X$ or a vertex $u\in V\setminus X$ of degree at least three, so:
\begin{equation}\label{eq:C}
C(v,X) = \min\{\min_{u\in X} \{C(u,X\setminus\{u\})+d(u,v)\}, \min_{u\in V\setminus X}\{B(u,X) + d(u,v)\} \}
\end{equation}
Where $d(u,v)$ is the shortest distance between $u$ and $v$. In order to avoid calculating
values $B(v,X)$ and $C(v,X)$ for the same states multiple times, we store the values
for all previously processed states in a map. Full C++ implementation can be found at \cite{WygosDrey2014}

\section{Iterative Rounding 1.39-approximation}
Our last implementation is the LP-based randomized 1.39-approximation algorithm
by Byrka {\it et al.}~\cite{byrka}. It is currently the best known approximation
algorithm and is based on the Iterative Rounding technique
introduced by Jain~\cite{jain}. In the Iterative Rounding method we solve
an LP-relaxation of the given problem, possibly obtaining a non-integer solution.
We then iteratively round some LP variables according to problem-specific rules
and resolve the modified LP, until we obtain an approximate solution to the original problem.

PAAL provides a generic framework for Iterative Rounding methods. Implementing an algorithm
within this framework is based on providing the following primitives (components):
\begin{itemize}
	\item \emph{Init} -- a functor responsible for initializing the LP for the given
		problem and initializing some additional data structures,
	\item \emph{DependentRound} -- a functor responsible for performing dependent
		LP rounding (rounding based on all of the LP variables values),
	\item \emph{SetSolution} -- a functor responsible for constructing the solution of
		the original problem.
	\item \emph{SolveLP} -- a functor responsible for solving the LP for the first time,
	\item \emph{ResolveLP} -- a functor responsible for resolving a previously solved and modified LP,
	\item \emph{StopCondition} -- a functor responsible for checking the stop condition for the
		Iterative Rounding main loop.
\end{itemize}

\subsection{1.39-approximation Algorithm}
The 1.39-approximation algorithm is based on an LP-relaxation known as
the directed-component cut relaxation (DCR). First we need to give some necessary definitions afte~\cite{byrka}.

Given a subset of terminals $T'\subseteq T$ and
a terminal $t\in T'$ we define a \emph{directed component} $C$ on terminals $T'$ with \emph{sink} $t$
as a minimum-cost Steiner tree on terminals $T'$, with edges directed towards $t$.
We call the terminals of a component $C$ other then $\textrm{sink}(C)$ as
$\textrm{sources}(C) = V(C)\cap T\setminus\{\textrm{sink}(C)\}$. We also denote the cost
of a component as $c(C)$, the set of all components as $C_n$ and we say that a component
$C$ \emph{crosses} a set $U\subseteq T$ if $C$ has at least one source in $U$ and the sink outside $U$.
By $\delta^{+}_{C_n}(U)$ we denote the set of directed components crossing $U$.

By selecting an arbitrary terminal $r$ as a root, we can now formulate the DCR:
\begin{equation}
\begin{aligned}
 & \mbox{minimize} & \sum_{C\in C_n} c(C)x_C & & \\
 & \mbox{such that} & \sum_{C\in\delta^{+}_{C_n}(U)}x_C\ge 1 & & \forall U\subseteq T\setminus\{r\}, U\neq\emptyset\\
 & & x_C \ge 0 & & \forall C\in C_n\\
\end{aligned}
\end{equation}

As the size of the set $C_n$ is exponential, we restrict it to a set $C_k$ of directed components
that contain at most $k$ terminals (where $k$ is a constant number). By replacing $C_n$ with $C_k$
in the DCR formulation, we obtain a $k$-DCR with polynomially many variables and exponentially many
constraints. Despite the exponential number of constraints, the $k$-DCR can be solved in polynomial
time using the so-called separation oracle (more details are given in the following sections).

Using the $k$-DCR formulation, we can give the pseudocode of the randomized 1.39-approximation
algorithm:

\begin{algorithm}
    \begin{algorithmic}
        \For{$i=1,2,\ldots$}
        \State $C_k \gets$ all components on at most $k$ terminals, each generated using
        Dreyfus-Wagner algorithm.
        \State Solve the $k$-DCR.
        \State Select one component $C_i$, where $C_i=C$ with probability $x_C/\sum_{C'\in C_k}x_{C'}$.
        \State Contract terminals of $C_i$ into its sink.
        \If{Only one terminal remains}
        \State $i_{max} \gets i$
        \State exit loop
        \EndIf
        \EndFor
        \State \Return $\bigcup_{i=1}^{i_{max}}C_i$
    \end{algorithmic}
    \caption{Pseudocode of the randomized $1.39$-approximation}
\end{algorithm}

\subsection{Algorithm Implementation}
To simplify the implementation we convert, without loss of generality, the input graph
into its metric closure (complete weighted graph on the same nodes, with weights given
by the shortest paths in the original graph).

The algorithm was implemented using the previously described PAAL Iterative Rounding
framework. The main part of the implementation are the necessary framework primitives:
\begin{itemize}
	\item \emph{steiner\_tree\_init} -- generates the $C_k$ set and initializes the LP.
		To generate $C_k$ we iterate over all subsets of $T$ of size at most $k$ and
		use the Dreyfus-Wagner algorithm to find the optimal Steiner tree on each subset.
	\item \emph{steiner\_tree\_round\_condition} -- selects one random component $C$ with
		probability $x_C/\sum_{C'\in C_k}x_{C'}$. Contracts terminals of $C$ into its sink
		and updates the metric distances from the contracted node and reinitializes $C_k$
		and the LP (using \emph{steiner\_tree\_init}).
	\item \emph{steiner\_tree\_stop\_condition} -- checks if the number of remaining
		terminals is equal to 1.
	\item \emph{steiner\_tree\_set\_solution} -- joins the sets of Steiner vertices from
		components selected in each phase.
\end{itemize}
The remaining primitives are the ones responsible for solving the LP. We detail their
implementation in the following section.

\subsubsection{Solving the LP}
As mentioned previously, the $k$-DCR LP has polynomially many variables but exponentially
many constraints. The authors of the original paper~\cite{byrka} show, that the $k$-DCR
can be reformulated into an equivalent polynomial sized LP by considering an nonsimultaneous
multicommodity flow problem in an auxiliary directed graph. Despite its polynomial size,
the equivalent formulation still has a large number of constraints: $O(k|T|^{k+1})$.
We modify the approach from~\cite{byrka} and provide a \emph{separation oracle} for the
$k$-DCR and use it together with the \emph{row generation} technique.

A \emph{separation oracle} for an LP is an algorithm, which given a solution of the LP
decides whether the solution is feasible or if not, returns a constraint violated
by the solution. We can use the separation oracle to implement the \emph{row generation}
technique. This technique uses the following approach: solve an LP that contains only
a subset of the constraints (the subproblem), let a basic optimal solution be $x_0$.
If the oracle shows that $x_0$ satisfies all the constraints, then $x_0$ is a basic optimal
solution of the original problem (since it is optimal for the subproblem, which is a relaxation,
and feasible for the original). If, on the other hand, the oracle finds a violated constraint,
then add this constraint to the subproblem, and iterate the process.

Let us now describe the separation oracle for the $k$-DCR. Consider an auxiliary directed
graph $G'=(V', E')$, where $V'=T\cup\{v_C | C\in C_k\}$. For every $C\in C_k$ we add an
edge $e_C=(v_C, \textrm{sink}(C))$ and edges $(u, v_C)$ for every $u\in\textrm{sources}(C)$.
Let the edges $e_C$ have weights $x_C$ (the value of variable $x_C$ in the solution being
checked by the oracle), while the other edges have infinite weights.

Now consider a minimum directed cut in $G'$ separating a vertex $v\in T\setminus\{r\}$ and $r$.
Let the cut be $(T_1\cup C_1, T_2\cup C_2)$, where $T_1,T_2\subset T$, $v\in T_1$, $r\in T_2$,
$C_1, C_2\subset\{v_C | C\in C_k\}$. If for some $u\in T_1$ there would exist a component $C_u$,
such that $u\in\textrm{sources}(C_u)$ and $v_{C_u}\in C_2$, then the weight of the cut would
be infinite and the cut would not be minimal. Thus for every $u\in T_1$ all components which
contain $u$ as a source must belong to $C_1$. The weight of the cut is equal to:

\begin{equation}
w(T_1\cup C_1, T_2\cup C_2)= \sum_{\{C\in C_1| \textrm{sink}(C)\in T_2\}}x_C
\end{equation}

It is easy to see, that for a given $T_1$ this weight is minimal when $C_1$ does not contain
any components $C'$, such that $\textrm{sources}(C')\cap T_1=\emptyset$. In such case the
weight of the cut is equal to the sum of $x_C$ for components, which have at least one source
in $T_1$ and the sink outside $T_1$ (components crossing $T_1$):

\begin{equation}
w(T_1\cup C_1, T_2\cup C_2)= \sum_{C\in \delta^{+}_{C_k}(T_1)}x_C
\end{equation}

That way, we can describe the separation oracle for $k$-DCR: for all $v\in T\setminus\{r\}$
we check (using a polynomial minimum cut algorithm) if the weight of the minimum directed cut
separating $v$ from $r$ is greater or equal to 1. If not, then the set $C_1$ defined by
the minimum cut gives us a violated constraint.

We tried several heuristics to improve the running time of the row generation.
First we tried to find the most violating constraint (that is, we iterate over all
$v\in T\setminus\{r\}$ and select the smallest of all found cuts). We also tried to stop
the search as soon as the first violated constraint was found (so we do not have to compute
all $|T|-1$ minimum cuts). The best running time was obtained by the following randomization:
we choose a random permutation of $T\setminus\{r\}$ every time we use the oracle (as opposed
to using the same permutation every time) and then we search until the first violated
constraint is found.

We need to note that the row generation algorithm does not have a polynomial running time
guarantee. An LP can be solved in polynomial time using a polynomial separation oracle
\cite{oracles} by the ellipsoid algorithm. However, because of the high complexity of the
ellipsoid algorithm the row generation method works better in practice.

\subsection{Components Generation}
Experiments with the implementation have shown that generation of components is the
bottleneck of the algorithm. To generate the set $C_k$, we iterate over all subsets
of $T$ with at most $k$ elements and for each of them we run the Dreyfus-Wagner algorithm.
Thus, the time complexity of this phase of the algorithm is $O(|T|^k\cdot(3^k|V|+2^k|V|^2))$.
Because of that, even for small values of $k$, for example $k=4$, this phase is the
bottleneck of the algorithm (for most instances component generation takes over 90\% or
even close to 100\% of the total runtime).

To improve the algorithm running time we tested some optimizations for the component generation
phase. First optimization comes from the fact, that we convert the graph into its metric closure.
In such complete graph, for every subset $T'$ of $T$ we can find a tree with leafs form $T'$ and
not containing any other terminals (a component on $T'$). However, for some subsets $T'$ there
may not exist such a tree in the original graph. Because of that,
we can ignore such subsets and decrease the number of calls to the Dreyfus-Wagner algorithm.
To decide if a subset $T'$ can produce a valid component, we need to check if there exists
a path in the original graph between each two terminals form $T'$ consisting only of non-terminals.
After initial preprocessing in $O(|T|\cdot(|V|+|E|))$ time (running a BFS algorithm from each
terminal), such check can be performed in $O(|T'|^2)$ time.

The speedup given by the above optimization depends heavily on the problem instance: it
gives an improvement only in cases where many terminal pairs cannot be connected by
a path consisting only of non-terminals. For such instances the optimization improved
the algorithm running time by up to 5 times, however the component generation still remained
the bottleneck of the implementation.

\subsubsection{Components Generation Optimization}
The performance of the component generation is determined by the time spend in the
Dreyfus-Wagner algorithm. In order to improve the component generation running times,
we need to look into the details of that algorithm.

The Dreyfus-Wagner method is based on recursive caltulation of functions $B(v,X)$, $C(v,X)$
for certain nodes $v\in V$ and terminal subsets $X\subseteq T$ (of decreasing size).
In the component generation phase of the IR algorithm, we repetitively run
the Dreyfus-Wagner algorithm for different sets of terminals. However, it is easy to
see, that functions $B$ and $C$ for many states $(v, X)$ are calculated for more then one
component. We can use this observation to implement the following optimization: we are
going to store the values of $B$ and $C$ for all previously calculated states (from all
previous components, not just the current one). Using the above optimization we were able
to reduce the time complexity of component generation from $O(\sum_{i = 2}^{k}{{{|T|}\choose{i}}(3^i|V| + 2^i|V|^2))}$ to
$O(\sum_{i = 2}^{k}{{{|T|}\choose{i}}(2^i|V| + |V|^2))}$.

This optimization gave a big improvement to the algorithm running time. For instances, which
previously were solved in under 10 minutes, the optimization gave an average speedup of
6-10 times (depending on $k$) and up to 100 times speedup for some instances.
It also increased the number of instances solved within the 10 minutes time limit
by 10\% for each tested parameter $k$. After the optimization, the
component generation phase took approximately 80\% of the total running time as opposed
to the previous 98-99\%, however it still remained the bottleneck of the algorithm.

\section{State-of-the-Art Local Search}
\label{sec:local-search}
We compare our approximation results with a results obtained using the best one out of the local search algorithms
proposed by Uchoa and Werneck in~\cite{wer}, i.e., multistart heuristic (MS). For completeness of the paper we we will describe shortly their
approach.

The multistart heuristic: works in two phases. During the first phase it builds
an initial MST using Shortest Path Heuristic (SPH). During the second phase it
improves the MST using Hill Climbing for as long as improvement are possible using three types of
moves. Authors restart the algorithm up to $100$ times using different starting points for the SPH.

\subsection{Shortest Path Heuristic}
Shortest Path Heuristic greedily builds a initial Steiner tree in as shown in Algorithm~\ref{alg-sph}.
\begin{algorithm}
    \begin{algorithmic}
        \State $tree \gets$ random\ node
        \While{there are some terminals not in $tree$}
        \State $t \gets$ terminal not in $tree$ that is closest to some node in $tree$
        \State $tree \gets tree +$ shortest path from $t$ to $tree$
        \EndWhile
    \end{algorithmic}
    \caption{Shortest Path Heuristic}
    \label{alg-sph}
\end{algorithm}
The tree built by the SPH depends on the choice of the starting node, and on choices of nodes and paths made
when there is a tie. Local Search algorithm builds only one tree for one starting
node, but the whole procedure is used multiple times for randomly selected starting nodes.

\subsection{Hill Climbing}

After an initial tree is build using the SPH, it is improved using three types of moves
via Hill Climbing method. The followind moves are applied to the tree until no further
improvement can be made.

\subsubsection{Steiner Node Insertion.}
The first move type is the insertion of a Steiner node into the tree.
We search for a vertex $v\notin tree$, such that the cost of
MST$(G[tree\cup \{v\}])$ is smaller than the cost of $MST(G[tree])$. If such node
is found it means we have found a tree spanning all terminals which has
a smaller cost than the previous one. We add this node to the set of
selected Steiner points.

\subsubsection{Key Path Exchange.}
A \emph{key node} in a Steiner tree is a non-terminal node with a degree at
least three. \emph{Crucial nodes} of a Steiner tree are all terminals
and key vertices. A \emph{key path} is a path that connects two crucial
nodes and has no internal crucial node.

The second kind of moves is a key path exchange. It removes a key path by splitting the tree
into two connected components and then reconnects them using a new path. All improving
key paths exchanges can be found by running Dijkstra's algorithm for each of
the $O(|T|)$ key paths in the tree. The total running time of this algorithm when using Fibonacci heaps would be
$O(|T|(|E| + |V|\log |V|))$ \cite{verhoeven1996local}.
This time can be improved by using sequences of Voronoi diagrams as shown in~\cite{wer}, what gives
an $O(|E| \log |V|)$ time algorithm for finding this improvement.

\subsubsection{Key Vertex Elimination.}
Key nodes succinctly describe Steiner tree of a graph~\cite{duin}. The third
type of moves is the key vertex elimination. Let $K$ be the set of key nodes in $tree$.
We want to find a vertex $v$ in $K$ such that the cost of MST$(G^*[K\cup T -v])$ is smaller
then MST$(G^*[K\cup T])$. Such vertex could be found by calculating MST for each of
possible $O(|T|)$ vertices $v\in K$, however, it would give us an
$O(|T|(|E| + |V|\log |V|))$ time algorithm~\cite{de2001hybrid}.
In~\cite{wer} authors improve running time of key vertex elimination to
$O(|E| \log |V|)$.

\section{Experiments}
\label{sec:exper}
We have tested our implementations on data sets from SteinLib~\cite{stlib}. All tests
were run with a timeout of $10$ minutes. Most of the algorithms did not manage to solve
all test cases in this time bound. The numbers of solved test cases are
shown in Table~\ref{tab:tes_count}. The table gives both the results for the IR
implementation with and without the component generation optimization. In our further
discussion we consider only the fastest (optimized) version of the algorithm.

\begin{table}
    \centering
    \caption{Numbers of solved SteinLib instances using different algorithms. IR optimized
    	refers to the IR implementation with the component generation optimization.}
\begin{tabular}{|l|r|r|}
        \hline
        Algorithm & Solved cases & Percent \\
        \hline
        \hline
        Dreyfus Wagner& 308& 30\%\\
        Greedy $2$ approx & 1021& 100\%\\
        Zelikovsky & 964& 94\%\\
        IR $k=2$& 770& 75\%\\
        IR $k=2$ optimized& 858& 84\%\\
		IR $k=3$& 695& 68\%\\
        IR $k=3$ optimized& 762& 75\%\\
		IR $k=4$& 550& 54\%\\
        IR $k=4$ optimized& 617& 60\%\\
		IR $k=5$& 389& 38\%\\
        IR $k=5$ optimized& 514& 50\%\\
        IR $k=6$ optimized& 452& 44\%\\
        \hline
    \end{tabular}
    \label{tab:tes_count}
\end{table}

All of our programs where compiled with \texttt{-O3} optimization option using
gcc 4.8.1. We used a $24$ core Intel Xeon CPU E5649@2.53GHz machine with Ubuntu
12.04 installed. The computer was equipped with 64GB of RAM.
For solving the LP in the Iterative Rounding algorithm we used the GLPK
LP solver \cite{glpk}. We have not implemented Local Search on our own, since we got results from
Uchoa and Werneck~\cite{wer}. Their running times are always lower than $3$ minutes and,
unlike us, they implemented the algorithm using C\#.


\subsection{Comparison of the Main Algorithms}
\begin{figure}[h]
\includegraphics[width=\textwidth]{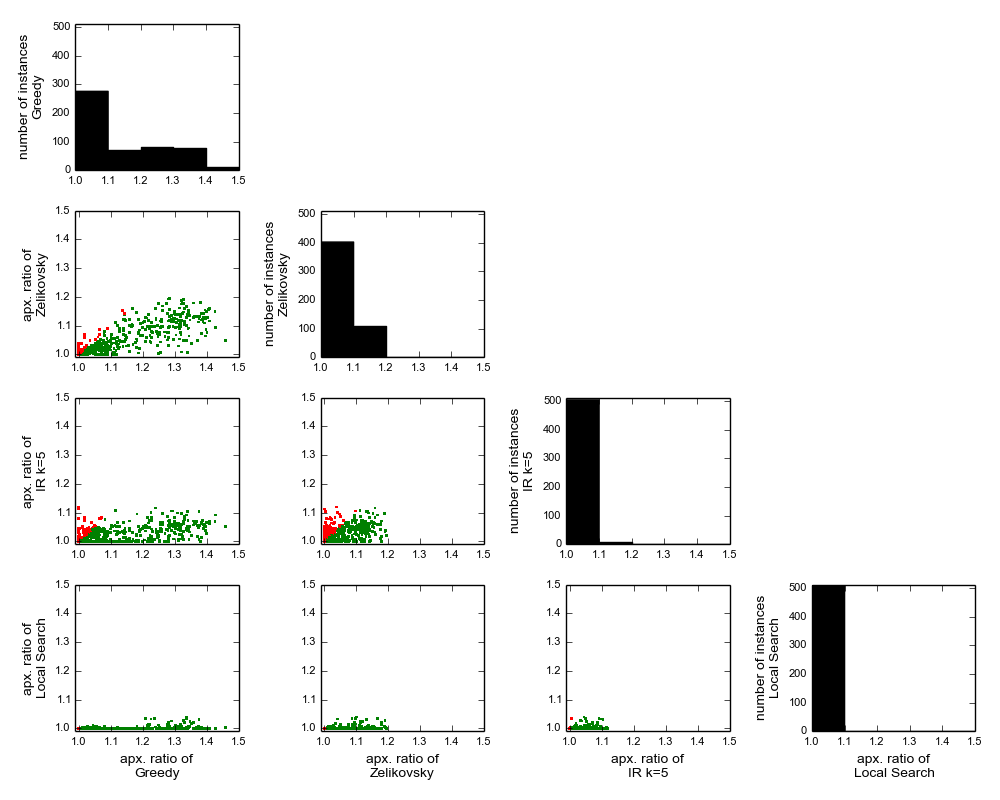}
\caption{Comparison of approximation ratios of following Steiner tree
    algorithms: Greedy, Zelikovsky, LS and IR ($k=5$) on cases from SteinLib.
    There are four histograms on the plot, one for each algorithm. Histograms' bins show
    number of test cases (on y-axis) that fell into each interval of approximation
    ratios (on x-axis). Scatter plots compare performance of pairs of algorithms.
    Points on scatter plots represent test cases with approximation ratios
achieved by compared algorithms on x and y-axis.}
\label{fig:cmpk5}
\end{figure}

Figure \ref{fig:cmpk5} shows comparison of approximation ratios of four main
algorithms: greedy 2-approximation, Zelikovsky algorithm, Local Search and IR for $k = 5$.
We define approximation ratio as cost of algorithm solution divided by best
known cost from SteinLib.
The figure shows only those cases for which all algorithms were able to find solutions
within the time limit, so there are $497$ points. There are four histograms and we
can easily see that for test cases from SteinLib Local Search achieves the best results,
being slightly better than Iterative Rounding.

Every point on each of scatter plots represents one test case with its approximation
ratios on $x$ and $y$ axis. We can see that there are cases for which
Iterative Rounding performs better than Local Search.

Table \ref{tab:cmpk5} shows average approximation ratios and running times in
seconds for each difficulty class, as defined in SteinLib \cite{stlib}.

\begin{table}
    \centering
\caption{Average approximation ratios and running times in seconds of our algorithms
for different SteinLib classes.}
\begin{tabular}{|l|rr|rr|rr|rr|r|}
    \hline
    Class &
    \multicolumn{2}{c|}{Greedy} &
    \multicolumn{2}{c|}{Zelikovsky} &
    \multicolumn{2}{c|}{IR $k = 5$} &
    \multicolumn{2}{c|}{IR $k = 5$} &
    Local Search \\
    &&&&&&& \multicolumn{2}{c|}{optimized} &\\

    & ratio & time & ratio & time & ratio & time & ratio & time & ratio\\
    \hline
	?? & 1.139 & 0.001 & 1.028 & 0.000 & 1.000 & 29.250 & 1.000 & 3.080 & 1.000\\
	?m & 1.338 & 0.161 & 1.133 & 5.815 & 1.040 & 105.350 & 1.040 & 11.395 & 1.002\\
	?s & 1.035 & 0.002 & 1.016 & 0.103 & 1.030 & 90.537 & 1.030 & 29.706 & 1.001\\
	Lm & 1.317 & 0.034 & 1.105 & 0.400 & 1.048 & 7.320 & 1.048 & 1.170 & 1.000\\
	Ls & 1.036 & 0.001 & 1.008 & 0.002 & 1.014 & 138.262 & 1.011 & 70.991 & 1.000\\
	NP? & 1.175 & 0.001 & 1.056 & 0.025 & 1.017 & 64.228 & 1.017 & 36.947 & 1.000\\
	NPm & 1.294 & 0.007 & 1.089 & 0.212 & 1.047 & 61.596 & 1.047 & 26.438 & 1.003\\
	NPs & 1.267 & 0.001 & 1.097 & 0.008 & 1.036 & 80.695 & 1.038 & 49.201 & 1.001\\
	Ph & 1.200 & 0.043 & 1.067 & 5.090 & 1.000 & 25.530 & 1.000 & 16.390 & 1.000\\
	Pm & 1.266 & 0.054 & 1.100 & 1.563 & 1.033 & 75.500 & 1.033 & 8.038 & 1.005\\
	Ps & 1.188 & 0.002 & 1.071 & 0.030 & 1.029 & 46.461 & 1.028 & 18.692 & 1.002\\

    \hline
	Average & 1.152 & 0.005 & 1.057 & 0.161 & 1.030 & 67.677 & 1.029 & 26.490 & 1.001\\
    \hline
\end{tabular}
\label{tab:cmpk5}
\end{table}

\subsection{Approximation Ratio of Iterative Rounding Depending on $k$}

\begin{figure}[h]
\includegraphics[width=\textwidth]{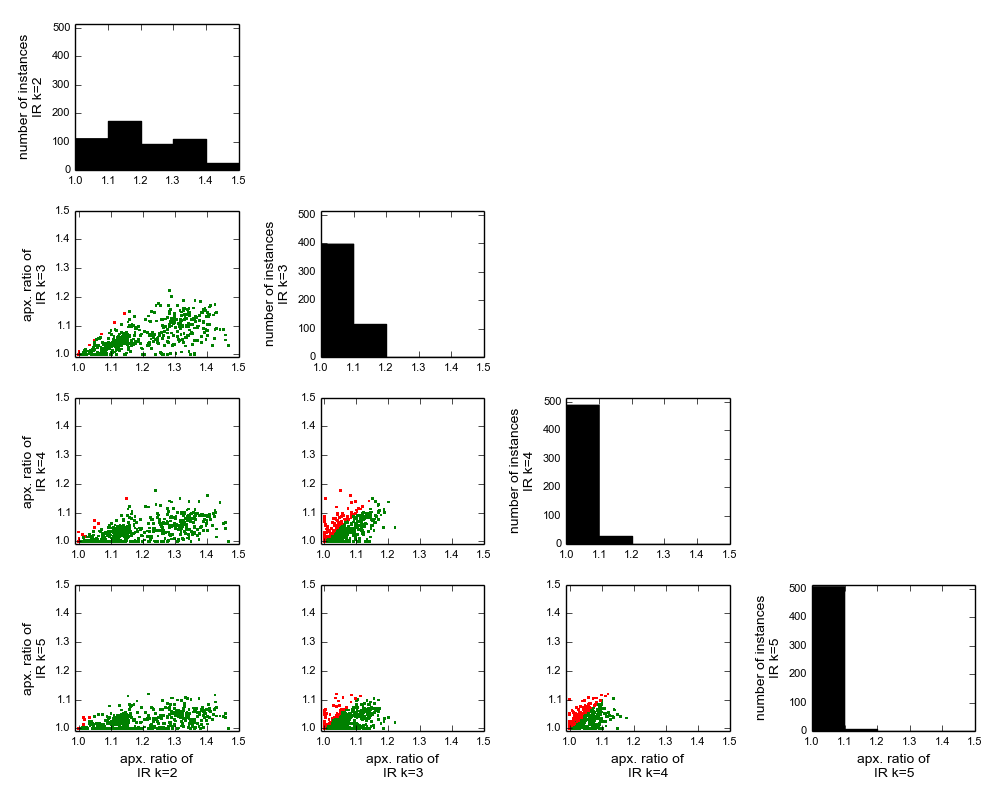}
\caption{Comparison of approximation ratios of Iterative Rounding depending
    on different values of $k$ on cases from SteinLib. There are four histograms
    on the plot, one for each value of $k\in \{2,3,4,5\}$. Histograms' bins show
    number of test cases (on y-axis) that fell into each interval of approximation
    ratios (on x-axis). Scatter plots compare performance of pairs of algorithms.
    Points on scatter plots represent test cases with approximation ratios
achieved for compared values of $k$ on x and y-axis.}
\label{fig:ir}
\end{figure}

Figure \ref{fig:ir} shows approximation ratios of IR depending
on the value of $k$. On average, with increasing $k$ approximation ratio
gets closer to $1$. There are some cases, where for bigger $k$ IR gives
worse results (we need to remember, however, that the IR is a randomized
approximation algorithm). Table \ref{tab:ir} shows average approximation ratios
and average running times of Iterative Rounding for different values of $k$
and different SteinLib dificulty classes.

\begin{table}
    \centering
\caption{Average approximation ratios and running times in seconds of the $1.39$-approximation
algorithm for different SteinLib classes and different values of $k$.}
\begin{tabular}{|l|rr|rr|rr|rr|}
    \hline
    Class &
    \multicolumn{2}{c|}{IR $k = 2$} &
    \multicolumn{2}{c|}{IR $k = 3$} &
    \multicolumn{2}{c|}{IR $k = 4$} &
    \multicolumn{2}{c|}{IR $k = 5$} \\
    & ratio & time & ratio & time & ratio & time & ratio & time  \\
    \hline
?? & 1.361 & 0.020 & 1.000 & 0.090 & 1.028 & 2.290 & 1.000 & 3.080\\
?m & 1.338 & 6.945 & 1.132 & 7.100 & 1.084 & 8.805 & 1.040 & 11.395\\
?s & 1.126 & 7.537 & 1.044 & 10.317 & 1.033 & 28.813 & 1.030 & 75.684\\
Lm & 1.317 & 0.440 & 1.108 & 0.460 & 1.055 & 0.920 & 1.048 & 1.170\\
Ls & 1.083 & 0.333 & 1.024 & 0.667 & 1.011 & 10.161 & 1.008 & 57.596\\
NP? & 1.371 & 1.125 & 1.045 & 16.150 & 1.029 & 26.250 & 1.017 & 36.947\\
NPh & 1.321 & 0.240 & 1.134 & 0.570 & 1.095 & 5.800 & 1.098 & 38.010\\
NPm & 1.353 & 0.243 & 1.095 & 0.515 & 1.061 & 3.909 & 1.051 & 31.811\\
NPs & 1.282 & 0.085 & 1.103 & 0.318 & 1.053 & 4.280 & 1.040 & 50.315\\
Ph & 1.213 & 9.810 & 1.060 & 8.770 & 1.060 & 40.855 & 1.020 & 122.940\\
Pm & 1.311 & 1.968 & 1.123 & 1.954 & 1.068 & 8.530 & 1.040 & 25.519\\
Ps & 1.233 & 0.356 & 1.068 & 0.464 & 1.046 & 6.172 & 1.027 & 34.033\\
    \hline
Average & 1.202 & 2.211 & 1.063 & 3.111 & 1.041 & 12.307 & 1.028 & 47.527\\
    \hline
\end{tabular}
\label{tab:ir}
\end{table}
\vspace{-0.3cm}

\subsection{Results}
Comparing the 3 approximation algorithms implemented as a part of this work
we see that for big enough parameters $k$ ($k\ge 4$), the results returned
by the IR algorithm are better then those returned by both the greedy
and Zelikovsky algorithms.

On the other hand, the running times of the IR algorithm are much higher then for
the other two algorithms. Also, while both the greedy and Zelikovsky algorithms were
able to solve over 90\% of SteinLib instances within our 10 minute time limit, the
IR for $k=4$ solved only 60\% of the instances, and that number decreases for bigger
values of $k$.

The MS Local Search heuristics from~\cite{wer} gives, on average, lower costs then all
of the algorithms with theoretical guarantees we have implemented. It was also able
to solve all SteinLib instances in under 3 minutes. Note, however, that the
paper~\cite{wer} gives 12 different versions of the local search and only the best
one of them visibly outperforms our IR algorithm. The other ones give worse
or comparable results.

Additionally, we have compared our IR results with few other papers. Our algorithm for $k>4$ gives
better results then the Tabu Search from~\cite{bastos2002reactive}, whereas it improves slightly over
the results from~\cite{de2002implementation} only on incidence class of SteinLib instances.

To summarize, the Iterative Rounding 1.39-approximation algorithm, which was the main interest of
this paper, does seem to give approximation results that compare decently with
other approaches. It is only outperformed by the best of Local Search implementations.
However, despite good theoretical approximation ratio and decent experimental quality of solutions, it's
running time is visibly higher then the one of the state-of-the-art heuristics like Local Search algorithms.
Nevertheless, the most important goal of this study, i.e., to demonstrate that high-level approximation
algorithm concepts can be implemented efficiently was accomplished successfully. Having implementations
of these concepts available it is easier to continue the work on hybrid solutions that would combine the best
aspects of different approaches and could potentially lead to better results.

\section{Acknowledgements}
We would like to thank Jarek Byrka for helpful comments on our optimization of components generation.

\bibliography{paper}
    \bibliographystyle{plain}
\end{document}